\documentclass[9pt,journal,compsoc]{IEEEtran}
\usepackage[labelfont={normalsize,sf,bf}, textfont=md]{caption}
\usepackage{fancyhdr}
\usepackage[normalem]{ulem}
\usepackage[sort,nocompress]{cite}
\usepackage[keeplastbox]{flushend}
\usepackage{microtype}
\usepackage{algorithm}
\usepackage[noend]{algopseudocode}
\usepackage{listings}
\usepackage{enumitem}
\usepackage{graphicx}
\usepackage{xspace}
\usepackage{subcaption}
\usepackage{makecell}
\usepackage[10pt]{moresize}
\DeclareGraphicsExtensions{.pdf,.jpg,.png}
\graphicspath{{./Figures/} {./figs/} {./plots/}{./Results/}}

\usepackage{amsmath}

\algblock{ParFor}{EndParFor}
\algnewcommand\algorithmicparfor{\textbf{par\_for}}
\algnewcommand\algorithmicpardo{\textbf{do}}
\algnewcommand\algorithmicendparfor{\textbf{end\ for}}
\algrenewtext{ParFor}[1]{\algorithmicparfor\ #1\ \algorithmicpardo}
\algrenewtext{EndParFor}{\algorithmicendparfor}
\algtext*{EndParFor}

\usepackage{multirow}

\newcolumntype{P}[1]{>{}p{#1\columnwidth}<{}}

\newcommand{\sys}{COBRA\xspace}

\newcommand{\pr}{{\tt Pagerank}\xspace}

\newcommand{\radii}{{\tt Radii}\xspace}

\newcommand{\PB}{PB\xspace}
\newcommand{\PBfull}{Propagation Blocking\xspace}

\newcommand{\neighpop}{{\tt Neighbor-Populate}\xspace}

\newcommand{\binning}{{\em Binning}\xspace}
\newcommand{\binread}{{\em Bin-Read}\xspace}
\newcommand{\coalbuf}{{\em C-Buffer}\xspace}
\newcommand{\coalbufs}{{\em C-Buffers}\xspace}

\newcommand{\binupd}{{\tt binupdate}\xspace}

\ifCLASSOPTIONcompsoc
  \usepackage[nocompress]{cite}
\else
  \usepackage{cite}
\fi
\ifCLASSINFOpdf
\else
\fi
\begin{document}
%
\title{Optimizing Graph Processing and Preprocessing with Hardware Assisted Propagation Blocking} 
\author{
    Vignesh Balaji \hspace{3em} Brandon Lucia \\
    Carnegie Mellon University \\
    {\em \{vigneshb, blucia\}@andrew.cmu.edu}
}
\markboth{}%
{}

\IEEEtitleabstractindextext{%
\begin{abstract}

Extensive prior research has focused on alleviating the characteristic poor 
cache locality of graph analytics workloads.
However, graph pre-processing tasks remain relatively unexplored.
In many important scenarios, graph pre-processing tasks can be as expensive
as the downstream graph analytics kernel.
We observe that Propagation Blocking (PB), a software 
optimization designed for SpMV kernels, generalizes to many graph 
analytics kernels as well as common pre-processing tasks.
In this work, we identify the lingering inefficiencies of a PB execution on
conventional multicores and propose architecture support to
eliminate PB's bottlenecks, further improving the performance gains from PB.
Our proposed architecture -- \sys -- optimizes the \PB execution of both 
graph processing and pre-processing alike to provide end-to-end speedups of up to 4.6x (3.5x on average).
\end{abstract}
}

\maketitle

\IEEEdisplaynontitleabstractindextext

%
\IEEEpeerreviewmaketitle

\IEEEraisesectionheading{\section{Introduction}\label{sec:introduction}}


The increasing main memory capacities and core counts of modern processors has motivated a
shift away from distributed graph processing towards analyzing graphs using just a single
machine~\cite{cost,ligra,gap}.
However, achieving high performance in single machine graph processing is challenging.
Irregular memory accesses lead to poor cache locality and an execution time dominated by
DRAM latency~\cite{cagra, navigating-maze}.
Considerable focus has been directed towards improving the performance of graph analytics
kernels using a variety of techniques including tiled executions, optimized data layouts, 
programming models, and custom architectures~\cite{cagra,graphgrind,gpop,pr-pcp,gorder,when-is-reordering-opt,rabbit-order,xstream,log-graph,slim-graph,droplet,graphicionado,isca2016-graph-accelerator,radar,graphit,phi,minnow,ozdal-pb,gridgraph,polymer,ligra+,galois}.

Most prior works focus solely on optimizing the graph analytics kernel execution time, assuming
an already pre-processed input graph stored in an optimized, sparse representation.
{\em Graph pre-processing} tasks that build the sparse graph representations are assumed to be 
amortized over many executions and, hence, have received little attention.
However, in scenarios such as processing time-evolving graphs~\cite{chronos,kineograph,graph-evolution} and 
one-shot processing~\cite{hats}, pre-processing times cannot be easily ignored~\cite{everything-afraid-graphs}.

In this work, we focus on the recently proposed {\em Propagation Blocking} (\PB) optimization~\cite{propagation-blocking}.
While \PB was developed as a software-based cache locality optimization for PageRank,
recent works have demonstrated \PB's effectiveness across a range of graph kernels~\cite{phi,milk}.
We observe that the key insight of PB -- exploiting unordered parallelism to 
{\em reorder} irregular updates for improved cache locality -- applies equally well to graph 
pre-processing.
The broad applicability of \PB across graph processing and pre-processing tasks makes \PB an ideal
candidate for acceleration with custom architecture support.

While \PB is an effective locality optimization~\cite{propagation-blocking}, all 
software-based \PB executions suffer from two fundamental inefficiencies.
First, we observe that \PB performance is sensitive to a software parameter called {\em bin range} (the
optimal value of which varies by application, input, and architecture).
Second, to reorder irregular memory accesses, \PB imposes non-trivial overheads in the form of executing many 
additional instructions.

We propose \sys: a set of modifications to the ISA and the memory hierarchy of a multicore processor to 
eliminate the inefficiencies of a \PB execution.
Instead of executing additional instructions to reorder irregular updates as in software-based \PB, \sys
introduces simple fixed-function logic in each level's cache controllers to efficiently reorder irregular updates.
\sys introduces a single (CISC-like) instruction to offload the reordering computation to the fixed function
units in each cache level.
\sys's architecture extensions eliminate the instruction overhead of \PB and also remove \PB's brittle dependence 
on the bin range parameter, instead defining \PB parameters by architectural properties
(size of each cache level).
We evaluate \sys across a common graph pre-processing task (converting an Edgelist representation of a graph
into a CSR) and a common graph processing kernel (\pr) and show that \sys provides a mean speedup 
of 1.74x over an optimized software \PB implementation.
Furthermore, since \sys applies to both graph processing and pre-processing, we show that \sys 
is able to provide an end-to-end speedup of 3.5x on average.

\section{Background}

The goal of this work is to improve the performance of graph processing 
{\em and} pre-processing.  
%
%
To provide background, we characterize the cost of graph (pre-)processing and 
provide an overview of the \PBfull (\PB) optimization that applies
to both graph processing and pre-processing.

\begin{figure}[h]
\centering
\includegraphics[keepaspectratio, width=0.90\columnwidth]{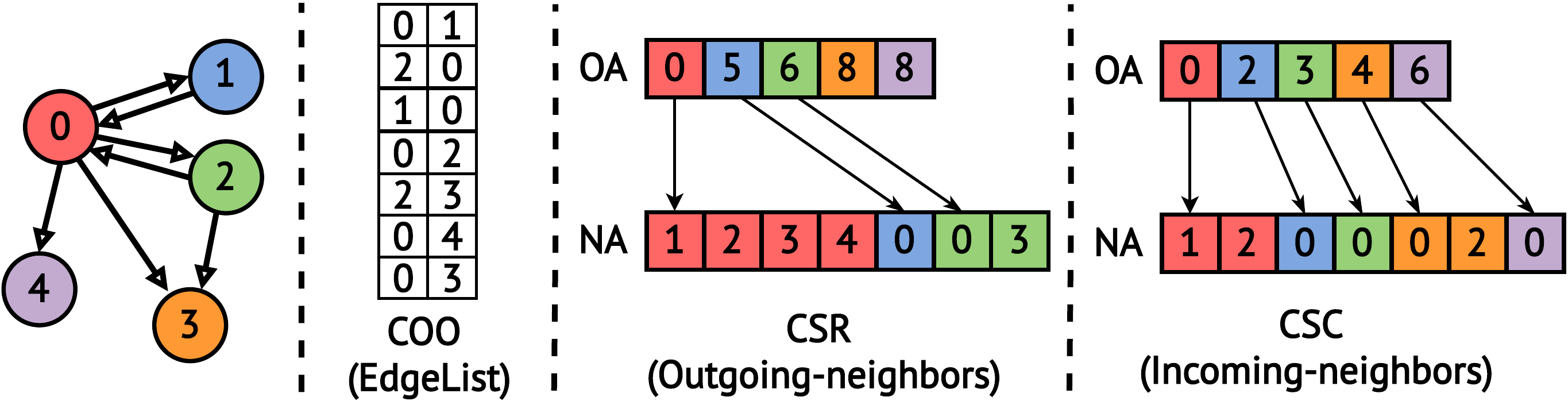}
\caption{{\bf Different representations of a directed graph}}
\label{fig:csr}
\end{figure}

{\noindent \bf Overview of Graph Processing:}
%
%
%
Graph analytics requires compressed representations because 
graphs are often extremely sparse (a typical adjacency matrix is $\geq$99\%
sparse~\cite{florida-sparse-matrices}).
%
%
Most frameworks~\cite{gap,ligra,graphit} use the {\em Compressed Sparse
Row} (CSR) format due to its memory efficiency and ability to quickly identify
a vertex's neighbors.
Figure~\ref{fig:csr} shows a graph's {\em outgoing} edges in CSR format.
CSR uses two arrays to represent edges, sorting by edge source IDs.
The Neighbor Array (NA) contiguously stores each vertex's neighbors.
The Offsets Array (OA) stores the starting offset of each vertex's neighbor
list in the NA.
%
%
%
Directed graphs use CSR for storing outgoing neighbors, and 
Compressed Sparse Column (CSC) to store the transpose of the adjacency matrix, which represents incoming neighbors.
%
%
%
Graph analytics kernels iteratively process vertices until meeting a convergence criterion. 
A graph kernel may iterate over a vertex's {\em outgoing} neighbors (``push'' execution) or 
{\em incoming} neighbors (``pull'' execution) or dynamically switch between the two neighborhoods~\cite{ligra,gap}.
Push-pull direction switching requires building both the CSC and CSR.

{\noindent \bf The Cost of Graph Pre-processing:\label{subsec:preprocessing-cost}} 
Most graph frameworks expect inputs in CSR/CSC format 
but graph repositories~\cite{florida-sparse-matrices} typically store 
graphs in the "coordinate list" COO format (also referred to as Edgelists)
as shown in Figure~\ref{fig:csr}.
%
%
Therefore, converting the input graph to CSR/CSC is a necessary 
pre-processing step before any useful computation happens.
%
The COO format is so prevalent that the Graph500 organization
has assigned the computation of converting an Edgelist into a different 
representation as one of three kernels used to benchmark 
supercomputers for their graph processing capabilities~\cite{introducing-graph500}.

\begin{figure}[h]
\centering
    \begin{subfigure}[b]{0.49\columnwidth}
            \includegraphics[keepaspectratio,width=0.96\columnwidth]{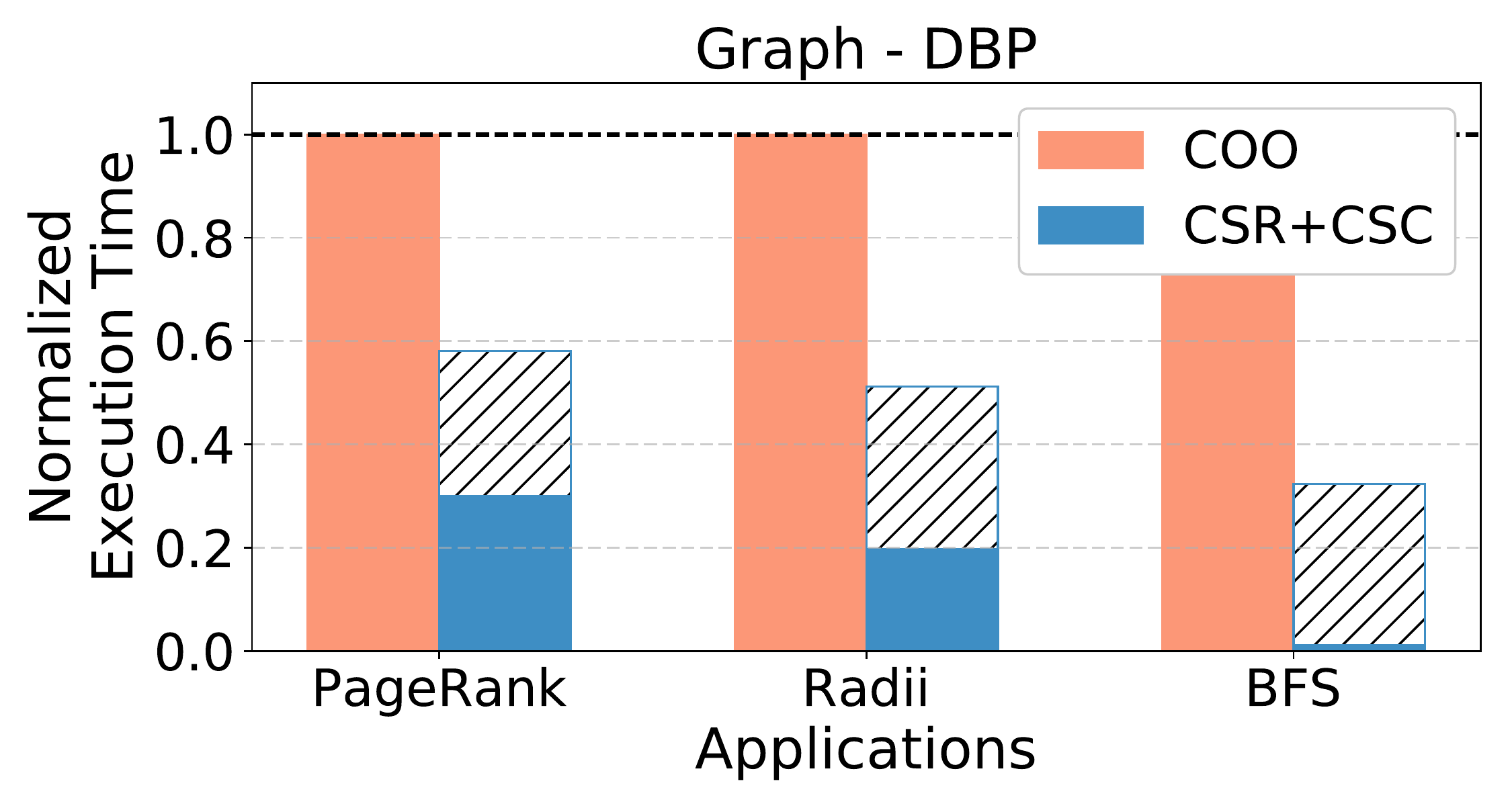}
            \caption{Edgelist to CSR}
            \label{subfig:el-to-csr-ovhd}
    \end{subfigure}%
    \begin{subfigure}[b]{0.49\columnwidth}
            \centering
            \includegraphics[keepaspectratio,width=0.96\columnwidth]{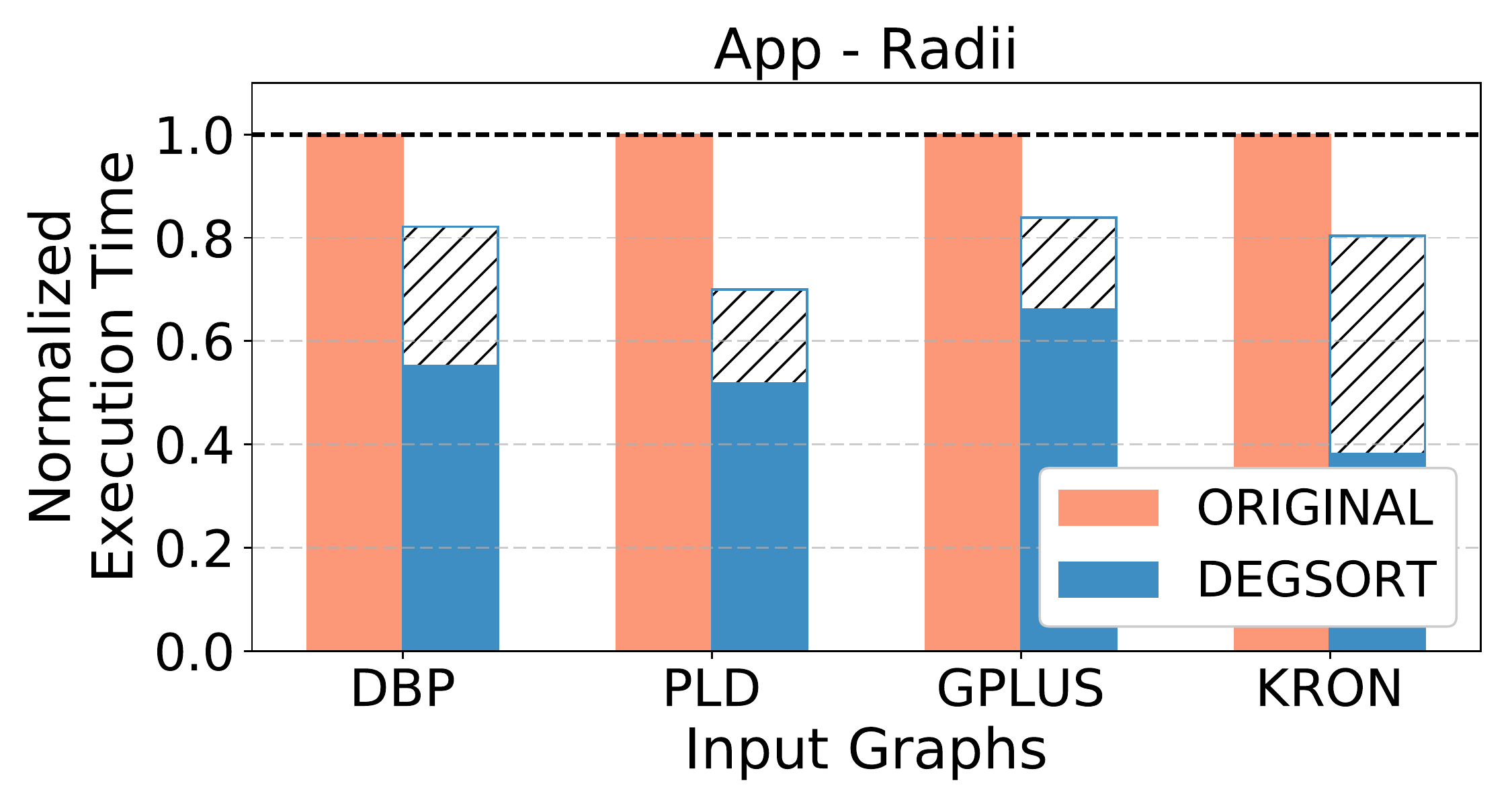}
            \caption{CSR to CSR}
            \label{subfig:lwr-ovhd}
    \end{subfigure}
    \caption{{\bf Pre-processing overheads:} 
        {\em (a) The shaded portion represents the cost of constructing a CSR (and CSC)
        from an Edgelist (COO) (b) The shaded portion represents the
        cost of creating a new CSR with the vertex order generated 
        by degree-sorting.
        }
    }
    \label{fig:preprocess-overheads}
\end{figure}

Graph pre-processing has a high cost.
Figure~\ref{subfig:el-to-csr-ovhd} compares the execution time of directly
processing an Edgelist compared to processing CSR/CSC.
The result shows the processing CSR data is faster even after
including the pre-processing cost to build the CSR. However, CSR
construction takes up 48\%-97\% of total execution time.
Constructing a CSR from an Edgelist (shown in 
Algorithm~\ref{algo:neighpop-unopt}) is not the only form of graph 
pre-processing.
{\em Graph reordering} is another popular pre-processing step where 
structural properties of graphs (e.g. community structure~\cite{phi,rabbit-order}) 
are exploited to change the labelling of vertex ids and create a new CSR with 
improved locality.
Figure~\ref{subfig:lwr-ovhd} shows the speedup of the \radii kernel running on
graphs produced by a lightweight reordering technique
(degree-sorting~\cite{when-is-reordering-opt}).  
The data show that lightweight reordering is effective even after including the cost
of constructing a new CSR.
However, constructing the reordered CSR takes up 25\%-55\% of the total run time.
The dominant computation in lightweight reordering and other sparse linear algebra 
pre-processing tasks such as constructing Compressed Sparse Fibers (CSF) for multi-dimensional 
tensors~\cite{splattsoftware} are variations of Algorithm~\ref{algo:neighpop-unopt}.
Therefore, we focus on the Edgelist-to-CSR kernel as a representative 
computation for graph pre-processing.

\begin{algorithm}
\caption{Kernel to populate neighbors (Edgelist-to-CSR)}
    \begin{algorithmic}[1]
             \State ${\tt offsets} \gets$ PrefixSum$({\tt degrees})$ \Comment{Offsets Array (OA)}
             \ParFor{e in $EL$}
                \State ${\tt neighs[offsets[e.src]]} \gets {\tt e.dst}$ \Comment{Neigh Array (NA)}
                \State AtomicAdd$({\tt offsets[e.src]}, 1)$
             \EndParFor
    \end{algorithmic}
\label{algo:neighpop-unopt}
\end{algorithm}

{\noindent \bf Propagation Blocking (PB):}
%
%
\PB improves cache locality of applications dominated by irregular memory
updates.
We explain the intuition behind \PB by applying it to the Edgelist-to-CSR
kernel shown in Algorithm~\ref{algo:neighpop-unopt} (henceforth
referred to as \neighpop).
A baseline execution of \neighpop suffers from poor cache locality because
edges may be arbitrarily ordered in the Edgelist, leading to fine-grain
irregular accesses to the {\tt offsets} array (Line 3; 
Algorithm~\ref{algo:neighpop-unopt}).
Accesses to the {\tt neighs} array are also irregular because they depend on 
the contents of the {\tt offsets} array.
A \PB execution of the \neighpop kernel (shown in Algorithm~\ref{algo:neighpop-pb})
improves locality by breaking the execution into two phases -- \binning
and \binread.
During the \binning phase, neither {\tt offsets} nor 
{\tt neighs} arrays are accessed.
Instead, pairs of indices and update values $(e.src, e.dst)$ are
stored in one of several {\em bins} maintained by \PB.
A bin is a data structure that {\em sequentially} stores each update 
belonging to a particular range (the "bin range") of data elements.
Once all updates have been written to bins, \PB starts the second phase -- \binread.
During \binread, the tuples (index and update value pairs) in a bin are sequentially processed
before moving to the next bin.
Since each bin stores updates for a smaller index range, the range of 
random accesses to the {\tt offsets} and {\tt neighs} arrays are reduced,
allowing each bin's updates to fit in on-chip caches.

\begin{algorithm}
\caption{PB version of Algorithm~\ref{algo:neighpop-unopt}}
    \begin{algorithmic}[1]
             \State ${\tt offsets} \gets$ PrefixSum$({\tt degrees})$ 
             \ParFor{e in $EL$}  \Comment{Binning Phase}
                \State ${\tt tid} \gets {\tt GetThreadID()}$
                \State ${\tt binID} \gets ({\tt e.src} / {\tt BinRange})$
                \State ${\tt bins[tid][binID]} \gets ({\tt e.src}, {\tt e.dst})$
             \EndParFor
             
             \ParFor{binID in NumBins}  \Comment{Bin-Read Phase}
                \For{tid in NumThreads}
                    \For {tuple in bins[tid][binID]}
                        \State ${\tt offsetVal} \gets {\tt offsets[tuple.src]}$
                        \State ${\tt neighs[offsetVal]} \gets {\tt tuple.dst}$
                        \State Add$({\tt offsets[tuple.src]}, 1)$
                    \EndFor
                \EndFor
             \EndParFor

    \end{algorithmic}
\label{algo:neighpop-pb}
\end{algorithm}

%
Prior works~\cite{phi,milk} have proposed optimizations to \PB for applications 
with {\em commutative updates}.
Commutativity allows coalescing multiple updates destined
to the same index, reducing the number of tuples to be written to bins 
which reduces total memory traffic.
We find that commutativity is {\em not necessary} to benefit from \PB.
The \neighpop kernel is an example of a non-commutative kernel. 
The updates to the {\tt offsets} array in \neighpop (Line 4; Algorithm~\ref{algo:neighpop-unopt}) are 
not commutative because the order of updates to the {\tt offsets} array determines the contents of the {\tt neighs} array (NA).
Coalescing updates to the {\tt offsets} array (as proposed in prior \PB optimizations~\cite{phi,milk}) would break 
correctness by skipping elements of the NA.
However, \PB on its own is still applicable to \neighpop because the kernel allows a vertex's neighbors to be 
listed in any order; the non-commutative updates permit {\em unordered parallelism}. 
Therefore, the applicability of \PB goes beyond just commutative updates (Table~\ref{table:pb-only-speedups}).

\begin{table}[h]
\centering
\scriptsize
\renewcommand{\tabcolsep}{2pt}
    \begin{tabular}{P{0.47}|P{.082}|P{.088}|P{.088}|P{.088}|P{.089}}
        \Xhline{1.0pt}
                                 & {\bf DBP} & {\bf KRON} & {\bf URND} & {\bf EURO} & {\bf HBUBL} \\
        \Xhline{1.0pt}
        {\bf NeighPop (Pre-processing) Speedup} & 6.9x  & 6.6x & 4.5x & 6.3x & 7.3x \\
        \hline
        {\bf PageRank (Processing) Speedup} & 1.3x & 1.1x & 1.2x & 0.8x & 1.2x \\  
        \hline
    \end{tabular}
\caption{\label{table:pb-only-speedups}{\bf Speedups from \PB}:
    {\em \PB is an effective optimization for both pre-processing and processing kernels across diverse input graphs}
}
\end{table}

\section{Opportunities for improving \PB}

We identify the opportunities for optimizing \PB without relying on update commutativity.
In this section, we characterize the two inefficiencies that all \PB executions on conventional multi-core
processors suffer from: (i) \PB must compromise by selecting a sub-optimal
{\em bin range} parameter, and (ii) binning updates in \PB requires executing many additional 
instructions.

\begin{figure}[h]
\centering
    
    \includegraphics[width=0.98\columnwidth]{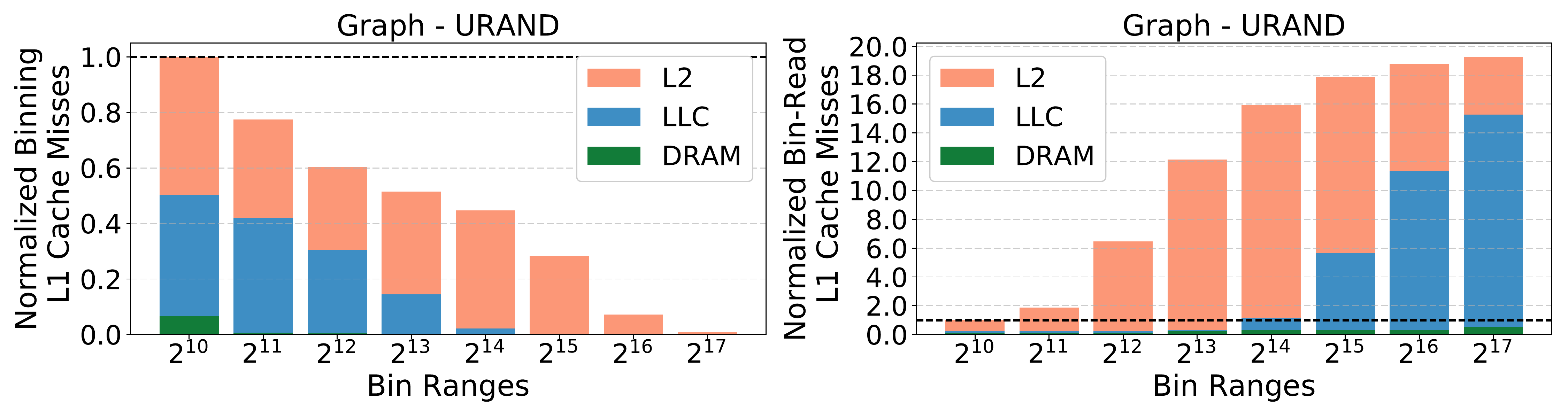}
    \caption{{\bf Sensitivity of PB to bin range:} 
        {\em Binning (left) prefers large bin ranges whereas Bin-Read (right) prefers small bin ranges.
        }
    }
    \label{subfig:perfctr-vs-binrange}
\end{figure}

{\bf Compromising on the Bin Range:}
The bin range parameter in \PB defines the range of indices mapped into a bin, and,
indirectly, the number of bins ($Bins = \frac{|Unique Indices|}{Bin Range}$).
To amortize the cost of writing to bins, the \binning phase uses cacheline-sized
{\em coalescing buffers} (henceforth referred to as \coalbufs) for each bin 
that accumulate updates to bins and enable coarse granularity writes to bins.
Consequently, the performance of both \binning and \binread is sensitive to the bin range.
Figure~\ref{subfig:perfctr-vs-binrange} shows performance counter results for 
\binning (left) and \binread (right) phases as bin range varies, for the \neighpop kernel. 
The data show normalized L1 load misses (broken into L2, LLC, and DRAM accesses).
\binning has the fewest cache misses with large bin ranges
because all the bins' \coalbufs fit in L1.
In contrast, \binread has the least cache misses with a small bin range 
because the range of indices modified by a bin's updates fits in the L1 cache (Lines 9-11; Algorithm~\ref{algo:neighpop-pb}).
Competing requirements on the bin range by the two phases forces all \PB 
executions to make a {\em compromise}, leading to 
sub-optimal performance in both phases (Table~\ref{table:pb-ideal-speedups}).

\begin{table}[h]
\centering
\scriptsize
\renewcommand{\tabcolsep}{2pt}
    \begin{tabular}{P{0.42}|P{.0982}|P{.0988}|P{.0988}|P{.0988}|P{.0989}}
        \Xhline{1.0pt}
                                 & {\bf DBP} & {\bf KRON} & {\bf URND} & {\bf EURO} & {\bf HBUBL} \\
        \Xhline{1.0pt}
        {\bf NeighPop + PB Speedup} & 6.9x  & 6.6x & 4.5x & 6.3x & 7.3x \\
        \hline
        {\bf NeighPop + PB-Ideal Speedup} & 9.5x  & 8.6x & 7.2x & 10.5x & 10.4x \\
        \hline
    \end{tabular}
\caption{\label{table:pb-ideal-speedups}{\bf Cost of compromising on the bin range}:
    {\em Allowing each of the two \PB phases to operate at their best bin ranges respectively (PB-Ideal) 
         can yield an additional (mean) speedup of 1.47x over PB} 
}
\end{table}

{\bf Overheads of binning in software \PB:}
The second major source of inefficiency in \PB is the need to execute additional instructions for binning updates and 
managing transfer of data between \coalbufs and bins in memory (Lines 4-5; Algorithm~\ref{algo:neighpop-pb}).
Executing additional instructions degrades instruction level parallelism by stressing core
resources (e.g., reservation stations, ROB, load-store queue).
Through simulation studies, we found that \PB executes up to 5x more instructions compared to a baseline
execution of \neighpop and causes the branch misprediction rate to increase from $\sim$0\% to $\sim$10\%.

Our characterization of \PB reveals two opportunities for improving \PB performance.
First, an efficient \PB execution must be able to avoid the compromise on the bin range parameter (i.e. simultaneously 
achieve optimal \binning and \binread performance).
Second, an efficient \PB execution must avoid the extra instruction overhead for binning.  


\section{\sys: An Architecture for efficient \PB} \label{sec:arch}

We present a new system called \sys\footnote{Since \PB is an instance of 
radix partitioning~\cite{propagation-blocking,radix-partitioning-surprising}, 
we named our system \sys (\underline{C}ache \underline{O}ptimized \underline{B}inning for 
\underline{Ra}dix Partitioning)} that specializes the cache hierarchy to 
eliminate the two inefficiencies of \PB executions.
\sys's architecture extensions are specifically targeted at improving \binning 
performance at small bin ranges.
Since \binread is naturally efficient at small bin ranges (Figure~\ref{subfig:perfctr-vs-binrange}),
the improved \binning performance allows \sys to achieve performance close to ideal \PB (Table~\ref{table:pb-ideal-speedups}).

\begin{figure}[h]
\centering
\includegraphics[keepaspectratio,width=0.96\columnwidth]{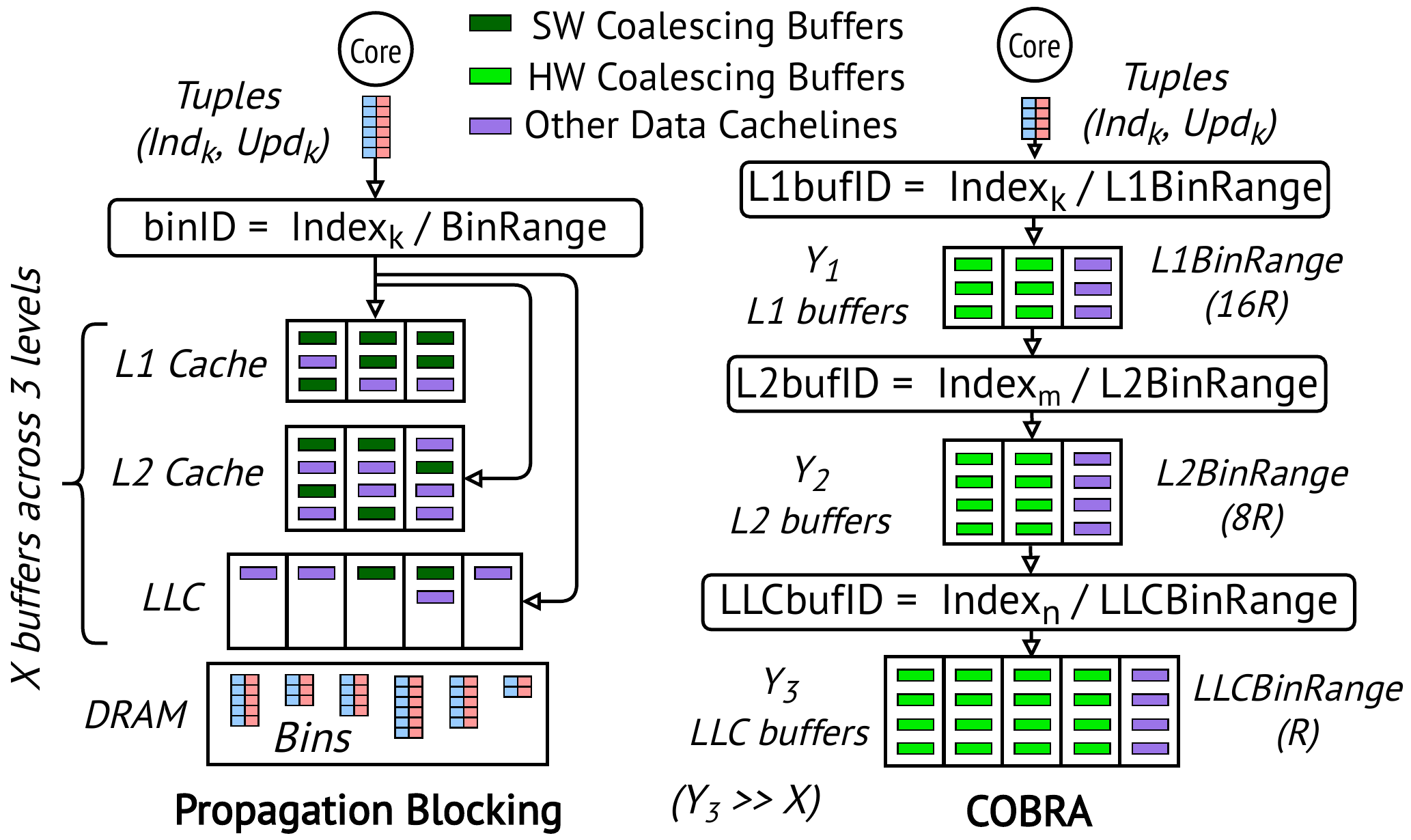}
\caption{{\bf Comparing the binning phases of \PB and \sys}: 
    {\em \sys maintains multiple levels of HW-managed C-Buffers to provide the illusion of large bin ranges for Binning while operating on small bin ranges for Bin-Read. We do not show bins in DRAM for \sys (\sys supports $Y_{3}$ bins
    in DRAM). The ratio of per-level bin ranges in \sys are dependent on the input and cache sizes.}
}
\label{fig:sys-high-level}
\end{figure}

{\bf Problem of Binning with small bin ranges:}
\binning has poor cache locality at small bin ranges.
Figure~\ref{fig:sys-high-level} (left part) explains why \binning in \PB performs poorly in a typical 3-level cache 
hierarchy.
At small bin ranges, the number of bins is high and all the \coalbufs do not fit in a small (e.g., L1) cache,
increasing the average latency of inserting tuples into \coalbufs.
Compounding the problem, increased cache demands by other program data displaces
\coalbufs to lower levels of cache (e.g., LLC) which further increases \coalbuf access latency.

{\bf Architecture support for Binning:}
The key insight of \sys is to {\bf decouple} \binning performance from the number of bins in memory.
Instead of a single bin range that spreads \coalbufs across the cache
hierarchy, \sys maintains a {\em hierarchy of C-Buffers}.
Each level of the cache hierarchy has a set of \coalbufs with  
a unique bin range that is specific to that level and maps tuples into
that level's \coalbufs.
The number of \coalbufs in a cache level is bounded by the capacity of that
level.
The L1 cache has the fewest \coalbufs (largest bin range) and the Last Level Cache (LLC) has the
most \coalbufs (smallest bin range).
For example, in Figure~\ref{fig:sys-high-level}, the L1, L2, and LLC each have $Y_{1}$, $Y_{2}$, and $Y_{3}$ 
\coalbufs with bin ranges $16R$, $8R$, and $R$ respectively.

\sys introduces a new instruction called \binupd whose operands are the index and 
update value that need to be binned.
The \binupd instruction interacts only with the L1 cache, writing tuples into one of the $Y_{1}$ \coalbufs 
identified by the {\tt L1BinRange} ($L1Buffer = \frac{Index}{16R}$).
When an L1 \coalbuf fills up with tuples, \sys does not transfer its contents 
directly to an in-memory bin (as in baseline \PB).
Instead, COBRA {\em evicts} the L1 \coalbuf by unpacking its tuples
and sending each tuple to its \coalbuf in the L2 cache.
Unlike a traditional cache eviction where the evicted line is sent
to the next cache level {\em as a whole}, during a \coalbuf eviction
each tuple in the filled \coalbuf may need to be written to a different \coalbuf
in the next cache level.
\sys writes each tuple evicted from the L1 \coalbuf into one of $Y_{2}$ \coalbufs
in the L2 cache identified by the {\tt L2BinRange} ($L2Buffer =
\frac{EvictedIndex}{8R}$).
Similarly, when an L2 \coalbuf fills up, \sys evicts it from L2 and sends each
of its tuples to one of the $Y_{3}$ \coalbufs present in the LLC.
Finally, when a LLC \coalbuf fills, \sys transfers all the tuples in the filled
LLC \coalbuf to the corresponding bin in main memory (as in baseline \PB).
In \sys, each \coalbuf eviction results in {\em scattering} tuples across the \coalbufs of the next cache
level. 
All tuples are first inserted into one of the L1 \coalbufs, then through evictions reach one of the LLC
\coalbufs, and finally (on LLC \coalbuf eviction) are written into one of $Y_{3}$ bins in memory.

The binning process in \sys is made efficient through two architecture extensions.
First, \sys relies on fixed-function units called {\tt binning engines} in each cache level's
controllers to handle \coalbuf evictions (including unpacking tuples from a filled $L_{i}$ \coalbuf and appending each tuple to an appropriate $L_{i+1}$ \coalbufs).
\sys uses simple way-based cache partitioning (as shown in Figure~\ref{fig:sys-high-level}) 
to pin \coalbufs to cache for the entirety of \binning, allowing simple logic to determine the unique location 
of a \coalbuf within a cache level.
The {\tt binning engines} allow \sys to offload \coalbuf management to hardware, eliminating the instruction
overhead of binning in \PB.
Second, \sys uses a small number of {\em eviction buffers} between cache levels to hide the latency 
of scattering tuples during \coalbuf evictions.
Removing the latency of \coalbuf evictions off the critical path is crucial in allowing \sys to achieve 
ideal \PB performance.
For the example in Figure~\ref{fig:sys-high-level}, the core sees a \binupd latency 
equivalent to binning into a small number of bins ($Y_{1}$) associated with a large bin range ($16R$) while actually operating on 
large number of bins ($Y_{3}$) associated with a small bin range ($R$).


\section{Experimental Methodology}

We characterized graph pre-processing costs (Figure~\ref{fig:preprocess-overheads}) 
and \PB's sensitivity to bin range (Figure~\ref{subfig:perfctr-vs-binrange})
on a real-system (an Intel Xeon processor with 14 cores and 35MB LLC).
For all other experiments, including evaluating \sys's performance, 
we use the Sniper~\cite{sniper} simulator to model an architecture with 
16 Out-of-Order (OoO) cores, a three-level cache hierarchy with 2MB/core NUCA LLC, 
and mesh interconnect.
We made various extensions to Sniper to model \sys -- adding support for non-temporal 
stores and ensuring that \binupd retires only when it reaches the head of the ROB 
because \binupd writes data caches (like stores).
We use baseline implementations of \neighpop and \pr from the GAP benchmark~\cite{gap}.
We evaluate each workload across large input graphs~\cite{florida-sparse-matrices} that are diverse in terms 
of degree-distribution (normal, power-law, bounded-degree), number of 
vertices (18-51 million), and average degrees (2-8).
All the input graphs far exceed the LLC capacity.
Finally, for the \PB runs, we use the original \PB source code which we received 
from the authors~\cite{propagation-blocking}.



\section{Evaluation}

\begin{figure}[h]
\centering
    \includegraphics[keepaspectratio,width=\columnwidth]{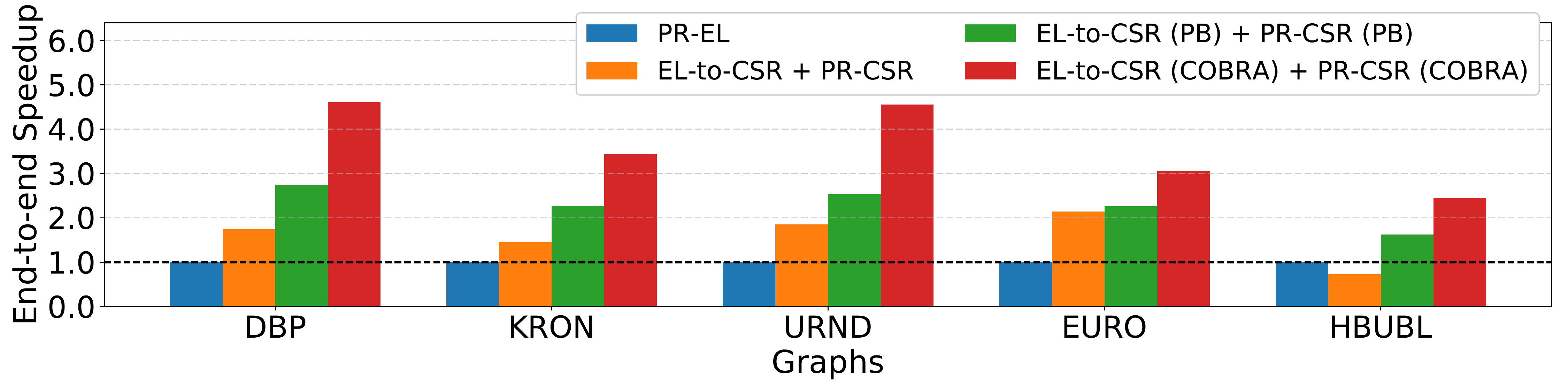}
    \caption{{\bf End to end speedups with \sys:} 
        {\em \sys applies to both graph pre-processing (EL-to-CSR) and processing (PageRank)}
    }
    \label{fig:end2end-speedups}
\end{figure}

\sys improves the performance of \PB substantially.
The main result of this evaluation is that \sys is effective across both graph processing and 
pre-processing.
We study the reduction in total execution time of the graph analytics
pipeline in Figure~\ref{subfig:el-to-csr-ovhd} -- building a CSR out of an 
Edgelist and running a graph analytics kernel on the CSR.
Figure~\ref{fig:end2end-speedups} shows the speedups of running \pr on an input graph initially stored
as an Edgelist.
The data show that even after including the pre-processing cost of constructing a CSR, running \pr on a CSR 
yields a mean speedup of 1.48x over running \pr directly on an Edgelist.
Applying \PB to both pre-processing and graph processing steps provides additional benefit, increasing the
mean speedup over Edgelist-based \pr to 2.25x.
Finally, \sys optimizes both \PB executions and increases the mean end-to-end speedup over 
Edgelist-based \pr to 3.5x.

\begin{figure}[h]
\centering
    \includegraphics[keepaspectratio,width=\columnwidth]{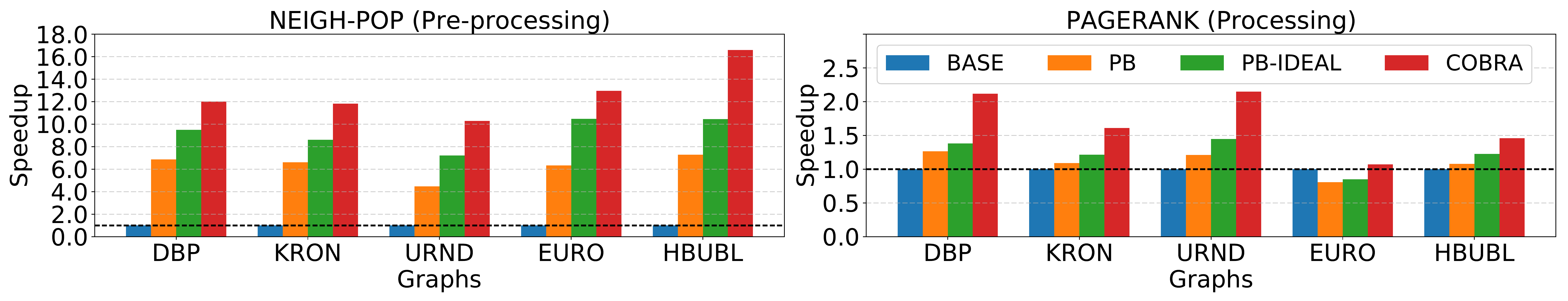}
    \caption{{\bf Breakup of \sys speedup} 
    }
    \label{fig:cobra-breakup}
\end{figure}

\sys improves \PB performance in two ways -- eliminating the compromise in bin range (achieving optimal 
\binning and \binread performance) and eliminating the instruction overheads associated with binning.
To isolate the contributions from each optimization, we compared the speedups from \PB, \PB-ideal (an 
idealized \PB execution combining \binning at a large bin range with \binread at a small bin range), and \sys. 
Figure~\ref{fig:cobra-breakup} shows that eliminating the compromise in bin range allows \PB-ideal to 
achieve a mean speedup of 1.28x over \PB.
As discussed in Section~\ref{sec:arch}, \sys achieves close to \PB-ideal performance by improving \binning
performance at small bin ranges.
Additionally, \sys reduces the instruction overheads of binning in \PB by offloading \coalbuf management to fixed-function
{\tt binning engines} in cache controllers.
By reducing the instruction overheads associated with binning, \sys is able to provide an additional speedup
of 1.35x over \PB-ideal.
Combining the benefits from eliminating the compromise on bin range and avoiding instruction
overheads of binning allows \sys to provide a mean speedup of 1.74x over \PB.

\section{Related Work}

%
Milk~\cite{milk} is a compiler-based approach that simplifies writing
\PB versions of applications. \sys could potentially be a backend
for the Milk compiler.
GraFBoost~\cite{grafboost} exploits commutativity in \PB for 
out-of-core graph analytics while \sys targets in-memory analytics.
PHI~\cite{phi} is hardware-based \PB optimization that uses simple ALUs at
each cache level to support in-place coalescing of updates.
Unfortunately, PHI relies on updates being commutative and is 
inapplicable to applications like \neighpop.
In contrast, \sys is applicable for non-commutative kernels and is a more
general optimization for \PB.
For commutative updates, we note that \sys could also employ simple ALUs within
caches as proposed in PHI and we are working towards incorporating PHI's 
update coalescing optimization into \sys.
%
%
%
Prior work~\cite{radix-partitioning-surprising, data-partitioning-myths} also highlighted \PB's sensitivity to bin range. 
\sys eliminates the need to tune the bin range by using a unique bin range for each cache level. 

\section{Conclusion}

We proposed \sys, a set of ISA extensions and cache hierarchy modifications, to accelerate \PB performance.
By eliminating the fundamental overheads of \PB executions, \sys achieves speedups of up to 2.3x 
over \PB.
As part of future work, we plan to investigate specializations for \sys based on 
application properties (e.g. commutativity) and expand \sys to other irregular application domains
beyond graph analytics.
\ifCLASSOPTIONcaptionsoff
  \newpage
\fi



\bibliographystyle{IEEEtran}
\bibliography{ref}

\begin{thebibliography}{10}
\providecommand{\url}[1]{#1}
\csname url@samestyle\endcsname
\providecommand{\newblock}{\relax}
\providecommand{\bibinfo}[2]{#2}
\providecommand{\BIBentrySTDinterwordspacing}{\spaceskip=0pt\relax}
\providecommand{\BIBentryALTinterwordstretchfactor}{4}
\providecommand{\BIBentryALTinterwordspacing}{\spaceskip=\fontdimen2\font plus
\BIBentryALTinterwordstretchfactor\fontdimen3\font minus
  \fontdimen4\font\relax}
\providecommand{\BIBforeignlanguage}[2]{{%
\expandafter\ifx\csname l@#1\endcsname\relax
\typeout{** WARNING: IEEEtran.bst: No hyphenation pattern has been}%
\typeout{** loaded for the language `#1'. Using the pattern for}%
\typeout{** the default language instead.}%
\else
\language=\csname l@#1\endcsname
\fi
#2}}
\providecommand{\BIBdecl}{\relax}
\BIBdecl

\bibitem{cost}
F.~McSherry, M.~Isard, and D.~G. Murray, ``Scalability! but at what cost?'' in
  \emph{HotOS}, 2015.

\bibitem{ligra}
J.~Shun and G.~E. Blelloch, ``Ligra: a lightweight graph processing framework
  for shared memory,'' in \emph{Proceedings of the 18th ACM SIGPLAN symposium
  on Principles and practice of parallel programming}, 2013, pp. 135--146.

\bibitem{gap}
S.~Beamer, K.~Asanovic, and D.~Patterson, ``Locality exists in graph
  processing: Workload characterization on an ivy bridge server,'' in
  \emph{Workload Characterization (IISWC), 2015 IEEE International Symposium
  on}.\hskip 1em plus 0.5em minus 0.4em\relax IEEE, 2015, pp. 56--65.

\bibitem{cagra}
Y.~Zhang, V.~Kiriansky, C.~Mendis, S.~Amarasinghe, and M.~Zaharia, ``Making
  caches work for graph analytics,'' in \emph{2017 IEEE International
  Conference on Big Data (Big Data)}, Dec 2017, pp. 293--302.

\bibitem{navigating-maze}
N.~Satish, N.~Sundaram, M.~M.~A. Patwary, J.~Seo, J.~Park, M.~A. Hassaan,
  S.~Sengupta, Z.~Yin, and P.~Dubey, ``Navigating the maze of graph analytics
  frameworks using massive graph datasets,'' in \emph{Proceedings of the 2014
  ACM SIGMOD international conference on Management of data}.\hskip 1em plus
  0.5em minus 0.4em\relax ACM, 2014, pp. 979--990.

\bibitem{graphgrind}
J.~Sun, H.~Vandierendonck, and D.~S. Nikolopoulos, ``Graphgrind: addressing
  load imbalance of graph partitioning,'' in \emph{Proceedings of the
  International Conference on Supercomputing}.\hskip 1em plus 0.5em minus
  0.4em\relax ACM, 2017, p.~16.

\bibitem{gpop}
\BIBentryALTinterwordspacing
K.~Lakhotia, R.~Kannan, S.~Pati, and V.~Prasanna, ``Gpop: A scalable cache- and
  memory-efficient framework for graph processing over parts,'' \emph{ACM
  Trans. Parallel Comput.}, vol.~7, no.~1, Mar. 2020. [Online]. Available:
  \url{https://doi.org/10.1145/3380942}
\BIBentrySTDinterwordspacing

\bibitem{pr-pcp}
K.~Lakhotia, R.~Kannan, and V.~Prasanna, ``Accelerating pagerank using
  partition-centric processing,'' in \emph{USENIX ATC}, 2018.

\bibitem{gorder}
H.~Wei, J.~X. Yu, C.~Lu, and X.~Lin, ``Speedup graph processing by graph
  ordering,'' in \emph{Proceedings of the 2016 International Conference on
  Management of Data}.\hskip 1em plus 0.5em minus 0.4em\relax ACM, 2016, pp.
  1813--1828.

\bibitem{when-is-reordering-opt}
V.~Balaji and B.~Lucia, ``When is graph reordering an optimization? studying
  the effect of lightweight graph reordering across applications and input
  graphs,'' in \emph{2018 IEEE International Symposium on Workload
  Characterization (IISWC)}.\hskip 1em plus 0.5em minus 0.4em\relax IEEE, 2018,
  pp. 203--214.

\bibitem{rabbit-order}
J.~Arai, H.~Shiokawa, T.~Yamamuro, M.~Onizuka, and S.~Iwamura, ``Rabbit order:
  Just-in-time parallel reordering for fast graph analysis,'' in \emph{Parallel
  and Distributed Processing Symposium, 2016 IEEE International}.\hskip 1em
  plus 0.5em minus 0.4em\relax IEEE, 2016, pp. 22--31.

\bibitem{xstream}
A.~Roy, I.~Mihailovic, and W.~Zwaenepoel, ``X-stream: Edge-centric graph
  processing using streaming partitions,'' in \emph{Proceedings of the
  Twenty-Fourth ACM Symposium on Operating Systems Principles}.\hskip 1em plus
  0.5em minus 0.4em\relax ACM, 2013, pp. 472--488.

\bibitem{log-graph}
M.~Besta, D.~Stanojevic, T.~Zivic, J.~Singh, M.~Hoerold, and T.~Hoefler, ``Log
  (graph) a near-optimal high-performance graph representation,'' in
  \emph{Proceedings of the 27th International Conference on Parallel
  Architectures and Compilation Techniques}, 2018, pp. 1--13.

\bibitem{slim-graph}
M.~Besta, S.~Weber, L.~Gianinazzi, R.~Gerstenberger, A.~Ivanov, Y.~Oltchik, and
  T.~Hoefler, ``Slim graph: practical lossy graph compression for approximate
  graph processing, storage, and analytics,'' in \emph{Proceedings of the
  International Conference for High Performance Computing, Networking, Storage
  and Analysis}, 2019, pp. 1--25.

\bibitem{droplet}
A.~Basak, S.~Li, X.~Hu, S.~M. Oh, X.~Xie, L.~Zhao, X.~Jiang, and Y.~Xie,
  ``Analysis and optimization of the memory hierarchy for graph processing
  workloads,'' in \emph{2019 IEEE International Symposium on High Performance
  Computer Architecture (HPCA)}.\hskip 1em plus 0.5em minus 0.4em\relax IEEE,
  2019, pp. 373--386.

\bibitem{graphicionado}
T.~J. Ham, L.~Wu, N.~Sundaram, N.~Satish, and M.~Martonosi, ``Graphicionado: A
  high-performance and energy-efficient accelerator for graph analytics,'' in
  \emph{Microarchitecture (MICRO), 2016 49th Annual IEEE/ACM International
  Symposium on}.\hskip 1em plus 0.5em minus 0.4em\relax IEEE, 2016, pp. 1--13.

\bibitem{isca2016-graph-accelerator}
M.~M. Ozdal, S.~Yesil, T.~Kim, A.~Ayupov, J.~Greth, S.~Burns, and O.~Ozturk,
  ``Energy efficient architecture for graph analytics accelerators,'' in
  \emph{2016 ACM/IEEE 43rd Annual International Symposium on Computer
  Architecture (ISCA)}.\hskip 1em plus 0.5em minus 0.4em\relax IEEE, 2016, pp.
  166--177.

\bibitem{radar}
V.~Balaji and B.~Lucia, ``Combining data duplication and graph reordering to
  accelerate parallel graph processing,'' in \emph{Proceedings of the 28th
  International Symposium on High-Performance Parallel and Distributed
  Computing}, 2019, pp. 133--144.

\bibitem{graphit}
Y.~Zhang, M.~Yang, R.~Baghdadi, S.~Kamil, J.~Shun, and S.~Amarasinghe,
  ``Graphit: a high-performance graph dsl,'' \emph{Proceedings of the ACM on
  Programming Languages}, vol.~2, no. OOPSLA, p. 121, 2018.

\bibitem{phi}
A.~Mukkara, N.~Beckmann, and D.~Sanchez, ``Phi: Architectural support for
  synchronization-and bandwidth-efficient commutative scatter updates,'' in
  \emph{Proceedings of the 52nd Annual IEEE/ACM International Symposium on
  Microarchitecture}, 2019, pp. 1009--1022.

\bibitem{minnow}
D.~Zhang, X.~Ma, M.~Thomson, and D.~Chiou, ``Minnow: Lightweight offload
  engines for worklist management and worklist-directed prefetching,'' in
  \emph{Proceedings of the Twenty-Third International Conference on
  Architectural Support for Programming Languages and Operating Systems}.\hskip
  1em plus 0.5em minus 0.4em\relax ACM, 2018, pp. 593--607.

\bibitem{ozdal-pb}
M.~M. {Ozdal}, ``Improving efficiency of parallel vertex-centric algorithms for
  irregular graphs,'' \emph{IEEE Transactions on Parallel and Distributed
  Systems}, vol.~30, no.~10, pp. 2265--2282, 2019.

\bibitem{gridgraph}
X.~Zhu, W.~Han, and W.~Chen, ``Gridgraph: Large-scale graph processing on a
  single machine using 2-level hierarchical partitioning.'' in \emph{USENIX
  Annual Technical Conference}, 2015, pp. 375--386.

\bibitem{polymer}
\BIBentryALTinterwordspacing
K.~Zhang, R.~Chen, and H.~Chen, ``Numa-aware graph-structured analytics,''
  \emph{SIGPLAN Not.}, vol.~50, no.~8, pp. 183--193, Jan. 2015. [Online].
  Available: \url{http://doi.acm.org/10.1145/2858788.2688507}
\BIBentrySTDinterwordspacing

\bibitem{ligra+}
J.~Shun, L.~Dhulipala, and G.~E. Blelloch, ``Smaller and faster: Parallel
  processing of compressed graphs with ligra+,'' in \emph{Data Compression
  Conference (DCC), 2015}.\hskip 1em plus 0.5em minus 0.4em\relax IEEE, 2015,
  pp. 403--412.

\bibitem{galois}
K.~Pingali, D.~Nguyen, M.~Kulkarni, M.~Burtscher, M.~A. Hassaan, R.~Kaleem,
  T.-H. Lee, A.~Lenharth, R.~Manevich, M.~M{\'e}ndez-Lojo \emph{et~al.}, ``The
  tao of parallelism in algorithms,'' in \emph{ACM Sigplan Notices}, vol.~46,
  no.~6.\hskip 1em plus 0.5em minus 0.4em\relax ACM, 2011, pp. 12--25.

\bibitem{chronos}
W.~Han, Y.~Miao, K.~Li, M.~Wu, F.~Yang, L.~Zhou, V.~Prabhakaran, W.~Chen, and
  E.~Chen, ``Chronos: a graph engine for temporal graph analysis,'' in
  \emph{Proceedings of the Ninth European Conference on Computer Systems},
  2014, pp. 1--14.

\bibitem{kineograph}
R.~Cheng, J.~Hong, A.~Kyrola, Y.~Miao, X.~Weng, M.~Wu, F.~Yang, L.~Zhou,
  F.~Zhao, and E.~Chen, ``Kineograph: taking the pulse of a fast-changing and
  connected world,'' in \emph{Proceedings of the 7th ACM european conference on
  Computer Systems}, 2012, pp. 85--98.

\bibitem{graph-evolution}
J.~Leskovec, J.~Kleinberg, and C.~Faloutsos, ``Graph evolution: Densification
  and shrinking diameters,'' \emph{ACM transactions on Knowledge Discovery from
  Data (TKDD)}, vol.~1, no.~1, pp. 2--es, 2007.

\bibitem{hats}
A.~Mukkara, N.~Beckmann, M.~Abeydeera, X.~Ma, and D.~Sanchez, ``{Exploiting
  Locality in Graph Analytics through Hardware-Accelerated Traversal
  Scheduling},'' in \emph{Proceedings of the 51st annual IEEE/ACM international
  symposium on Microarchitecture (MICRO-51)}, October 2018.

\bibitem{everything-afraid-graphs}
J.~Malicevic, B.~Lepers, and W.~Zwaenepoel, ``Everything you always wanted to
  know about multicore graph processing but were afraid to ask,'' in \emph{2017
  $\{$USENIX$\}$ Annual Technical Conference ($\{$USENIX$\}$$\{$ATC$\}$ 17)},
  2017, pp. 631--643.

\bibitem{propagation-blocking}
S.~Beamer, K.~Asanovi{\'c}, and D.~Patterson, ``Reducing pagerank communication
  via propagation blocking,'' in \emph{Parallel and Distributed Processing
  Symposium (IPDPS), 2017 IEEE International}.\hskip 1em plus 0.5em minus
  0.4em\relax IEEE, 2017, pp. 820--831.

\bibitem{milk}
\BIBentryALTinterwordspacing
V.~Kiriansky, Y.~Zhang, and S.~Amarasinghe, ``Optimizing indirect memory
  references with milk,'' in \emph{Proceedings of the 2016 International
  Conference on Parallel Architectures and Compilation}, ser. PACT ’16.\hskip
  1em plus 0.5em minus 0.4em\relax New York, NY, USA: Association for Computing
  Machinery, 2016, p. 299–312. [Online]. Available:
  \url{https://doi.org/10.1145/2967938.2967948}
\BIBentrySTDinterwordspacing

\bibitem{florida-sparse-matrices}
T.~A. Davis and Y.~Hu, ``The university of florida sparse matrix collection,''
  \emph{ACM TOMS}, 2011.

\bibitem{introducing-graph500}
R.~C. Murphy, K.~B. Wheeler, B.~W. Barrett, and J.~A. Ang, ``Introducing the
  graph 500,'' \emph{Cray Users Group (CUG)}, vol.~19, pp. 45--74, 2010.

\bibitem{splattsoftware}
S.~Smith and G.~Karypis, ``{SPLATT: The Surprisingly ParalleL spArse Tensor
  Toolkit},'' \url{http://cs.umn.edu/~splatt/}, 2016.

\bibitem{radix-partitioning-surprising}
F.~M. Schuhknecht, P.~Khanchandani, and J.~Dittrich, ``On the surprising
  difficulty of simple things: the case of radix partitioning,''
  \emph{Proceedings of the VLDB Endowment}, vol.~8, no.~9, pp. 934--937, 2015.

\bibitem{sniper}
\BIBentryALTinterwordspacing
T.~E. Carlson, W.~Heirman, and L.~Eeckhout, ``Sniper: Exploring the level of
  abstraction for scalable and accurate parallel multi-core simulation,'' in
  \emph{Proceedings of 2011 International Conference for High Performance
  Computing, Networking, Storage and Analysis}, ser. SC ’11.\hskip 1em plus
  0.5em minus 0.4em\relax New York, NY, USA: Association for Computing
  Machinery, 2011. [Online]. Available:
  \url{https://doi.org/10.1145/2063384.2063454}
\BIBentrySTDinterwordspacing

\bibitem{grafboost}
\BIBentryALTinterwordspacing
S.-W. Jun, A.~Wright, S.~Zhang, S.~Xu, and Arvind, ``Grafboost: Using
  accelerated flash storage for external graph analytics,'' in
  \emph{Proceedings of the 45th Annual International Symposium on Computer
  Architecture}, ser. ISCA ’18.\hskip 1em plus 0.5em minus 0.4em\relax IEEE
  Press, 2018, p. 411–424. [Online]. Available:
  \url{https://doi.org/10.1109/ISCA.2018.00042}
\BIBentrySTDinterwordspacing

\bibitem{data-partitioning-myths}
Z.~Zhang, H.~Deshmukh, and J.~M. Patel, ``Data partitioning for in-memory
  systems: Myths, challenges, and opportunities.'' in \emph{CIDR}, 2019.

\end{thebibliography}
\end{document}